\documentclass{article}
\usepackage{amsmath}
\usepackage[format=plain,
            font=it]{caption}
\usepackage{float}
\usepackage{hyperref}
\usepackage{cite}
\floatstyle{plaintop}
\restylefloat{table}

\title{Evidential strength of categorical expert witness statements}
\author{R.J.F.Ypma}

\begin{document}

\maketitle

In many jurisdictions, forensic evidence is presented in the form of categorical statements by forensic experts. 
In recent years, the scientific validity of forensic science practices have been questioned. 
In response, several large-scale performance studies have been performed, reporting various error rates for different forensic fields. 
This important work has elucidated the uncertainty associated with categorical statements.

There is growing scientific consensus that the likelihood ratio (LR) framework is the logically correct form of presentation for forensic evidence evaluation. Yet, the very relevant results from the large-scale performance studies have not been cast in this framework.

Here, I show how to straightforwardly calculate an LR for any given categorical statement using data from the performance studies.
This number quantifies how much more we should believe the hypothesis of same source vs different source, when provided a particular expert witness statement.
LRs are reported for categorical statements resulting from the analysis of latent fingerprints, bloodstain patterns, handwriting, footwear and firearms.
The highest LR found for statements of identification was 376 (fingerprints), the lowest found for statements of exclusion was 1/28 (handwriting).

The LRs found may be more insightful for those used to this framework than the various error rates reported previously. 
An additional advantage of using the LR in this way is the relative simplicity; there are no decisions necessary on what error rate to focus on or how to handle inconclusive statements.
The values found are closer to 1 than many would have expected. 
One possible explanation for this mismatch is that we undervalue numerical LRs.
Finally, a note of caution: the LR values reported here come from a simple calculation that does not do justice to the nuances of the large-scale studies and their differences to casework, and should be treated as  ball-park figures rather than definitive statements on the evidential value of whole forensic scientific fields.

\newpage

\section{Introduction}
Forensic evidence plays a large role in criminal law. In many jurisdictions, expert witnesses give categorical statements, i.e. claiming identification of a person or object from forensic traces. In recent decades forensic sciences have been criticized for ignoring or being unable to quantify the uncertainty associated with such statements \cite{pcast, council}. In response, several large-scale studies have been published that investigate this uncertainty for a range of forensic fields \cite{bloodstain, bullets, fingerprint, footwear, handwriting}. In these studies, uncertainty is quantified as error rates. This is however not a trivial computation, as several relevant error rates exist (e.g. false negative rate, false positive rate, false discovery rate) and there is no single correct way of handling statements of the form `unsuitable for analysis' or `inconclusive' \cite{Koehler2023}. 

The likelihood ratio (LR) framework provides an alternative form of reporting evidence, and is widely recommended as logically correct \cite{afsp, enfsi, asa, Koehler2023}. The likelihood ratio is defined as the likelihood of the evidence $E$ under one hypothesis ($H_1$) divided by the likelihood under a competing hypothesis ($H_2$):
$$
LR=P(E|H_1)/P(E|H_2)
$$
Adoption of numerical LRs is largest in the field of forensic DNA analysis, which boasts well-developed statistical models. 
However, an increasing number of forensic fields is following suit in developing numerical LR systems \cite{stijn}. 
The framework is equally applicable to subjective judgements, e.g. taking the form `\textit{the evidence provides strong support for the first proposition relative to the alternative}'.

No trier-of-fact calculates a probability of guilt. 
Rather, there is a psychological process that incorporates the evidence and results in a conviction of proven or not-proven. 
It is unclear whether this process takes into account uncertainty, expressed by error rates or likelihood ratios, in the way that the forensic scientist intended. Most likely, it does not \cite{Marquis2016-gg, Thompson2015, Sjerps1999, Eldridge2019, De_Keijser2012-bo}.
It seems reasonable that training in and exposure to particular forms of expressing evidence will improve interpretation \cite{Rassin2022}.
Thus, for those used to the likelihood ratio framework, it may be of use to quantify the uncertainty measured in the performance studies in the form of an LR.

Below, I will illustrate how likelihood ratios for expert witness statements can be straightforwardly computed from performance studies.
In fact, if only the two statements of `same source' or `different source' are used, the LR is simply (1 - false negative rate) / false positive rate. 
The interpretation of these calculated LRs is the support for same source relative to different source that is given by a particular statement, from an expert and sample pair comparable to those in the performance study. None of these numbers should be treated as a thorough assessment of the evidential strength of a statement in the context of a specific case.

\section{Methods}
We assume we have access to many evaluations made by a set of experts for which ground truth is known. Scott et al. \cite{bullets} provides such data for firearm examiners. Possible conclusions for the examiner to reach were individualisation (`ID'), three levels of inconclusive (`Inconcl. A-C'), `Elimination' and `Other'. For example, for the statement of individualisation the LR would be:
\begin{equation*}
\label{eq:LR}
LR = \frac{P(\textrm{expert concludes bullet fired from firearm}|\textrm{bullet was fired from firearm})}{P(\textrm{expert concludes bullet fired from firearm}|\textrm{bullet was not fired from firearm})}
\end{equation*} Table \ref{tab:bullets} shows how often conclusions were reached for matching and non-matching pairs. 

The LR can be calculated in a straightforward manner. The probability in the numerator is equal to the number of times that an expert said `ID' when $H_1$ was true, i.e. the bullet was fired from the firearm, divided by the total number of times $H_1$ was true. The first is 1076, the second is 1076+127+125+36+41+24 = 1429. This gives us a probability of 

\begin{equation*}
\begin{split}
 &    P(\textrm{expert concludes bullet fired from firearm}|\textrm{bullet was fired from firearm}) \\
& = 1076/1429 \\
& \approx  0.75
\end{split}
\end{equation*}

Likewise for the denominator, we get

\begin{equation*}
\begin{split}
& P(\textrm{expert concludes bullet fired from firearm}|\textrm{bullet was not fired from firearm}) \\
& = 20/2891 \\
& \approx  0.007
\end{split}
\end{equation*}

The ratio of these probabilities is the likelihood ratio: $LR \approx 0.75/0.007\approx109$. This calculation is exactly the same for all studies we looked at. Sometimes an additional tallying step is needed as the data are not given in aggregated form. Code is available from github\footnote{\url{github.com/NetherlandsForensicInstitute/evidential_value_of_expert_witness_statements}}.

\begin{table}[H]
    \centering
    \begin{tabular}{c|c|c|c|c|c|c}
    \hline
    & ID & Inconcl.-A & Inconcl.-B & Inconcl.-C & Elimination & Other\\ 
    \hline
    Matching & 1,076 & 127 & 125 & 36 & 41 & 24\\
    Non-matching & 20 & 268 & 848 & 745 & 961 & 49\\
    \hline
    \end{tabular}
    \caption{Statistics on bullet evaluations by expert witnesses, from \cite{bullets}.}
    \label{tab:bullets}
\end{table}

\newpage
\section{Results}
Table \ref{tab:bullet_results} gives the LRs for all categories considered for bullet evaluations. LRs for all different statement categories considered for the other performance studies (footwear, cartridge, handwriting, fingerprints, bloodstain patterns) are given in the tables in the appendix. Table \ref{tab:identification_results} summarises these by giving for every study the LR calculated for the identification and exclusion statements.

\begin{table}[H]
    \centering
    \begin{tabular}{c|c|c|c|c|c|c}
    \hline
    & ID & Inconcl.-A & Inconcl.-B & Inconcl.-C & Elimination & Other\\ \hline
    LR & 109 & 1 & 1 / 3 & 1 / 10 & 1 / 12& 1 \\
    \hline
    \end{tabular}
    \caption{Likelihood ratios for bullet evaluation conclusions by expert witnesses, calculated from data in \cite{bullets}.}
    \label{tab:bullet_results}
\end{table}

\begin{table}[H]
    \centering
    \begin{tabular}{l|c|c}
    \hline
    & LR  (identification) & LR (exclusion)\\
    \hline
    pattern type from bloodstain pattern \cite{bloodstain} & 6 & 1/4 \\ 
    writer from handwriting sample \cite{handwriting} & 17  & 1/28\\
    firearm from cartridge \cite{bullets} & 81 & 1/28  \\
    firearm from bullet \cite{bullets} & 109 & 1/12 \\
    footwear from print \cite{footwear}  & 113 & 1/7 \\
    person from latent fingerprints \cite{fingerprint} & 376 & 1/11\\  
    \hline
    
    \end{tabular}
    \caption{Likelihood ratios for expert witness from various fields, for statements of identification and exclusion.}
    \label{tab:identification_results}
\end{table}

\section{Discussion}
This study presents  the likelihood ratios for expert witness statements for five forensic fields from large performance studies.
The LRs are an alternative to error rates for quantifying the uncertainty associated with these statements.
This alternative may be useful to those who are more used to the LR framework than error rates, and may help bridge the gap between the two frameworks.
Incidentally, the LR framework simplifies the uncertainty analysis as we don't have to consider various different error rates and ways of handling `inconclusive' statements.

The likelihood ratios reported should be treated as `ballpark' figures, giving some insight into what LRs to associate with particular statements, such as identification.
We should be very hesitant in interpreting the LRs reported here as the evidential strength of an expert witness statement in a given case. 
The performance studies measure the performance of a set of experts, which may not be representative for the case, on a specific set of samples, which may not be representative for the case.
For example, several studies explicitly select different-source samples that are very challenging to evaluate \cite{handwriting, fingerprint, bullets}.
As an illustration, considering only the top 1\% hardest different-source comparisons would result in an LR 100 times lower (assuming no errors would be made on the remaining 99\%).
Furthermore, it is common practice in casework to have forensic witness evaluations reviewed by an independent expert - this extra quality check is absent in the performance studies.
Lastly, we stress that evidential strength is only one aspect of the value of forensic evidence; moderately strong evidence that a suspect held the murder weapon may be much more relevant than very strong evidence that he was near the crime scene.

The actual values of LRs found are interesting.
As expected, statements of increasing certainty result in higher LRs and `inconclusive' statements generally result in LRs close to 1.
More surprising is that even the highest LRs found for inclusion were below 400, and the lowest found for exclusion above 1/50.
This range is small. 
For example, both experts and the public expect a statement of fingerprint identification to be stronger than an LR of 100,000 \cite{Busey2023}.
The caveats mentioned above only partly explain the difference of several orders of magnitude between expected and calculated LRs.
It seems we either overvalue categorical statements, undervalue numerical LRs, or both.
This undervaluation is something we anecdotally encounter in practice, and is perhaps partly caused by the forensic science community. 
After all, an LR of 1000 is strong enough to convert a weak prior probability of 0.10 to a posterior of 0.99. 
Indeed, an LR of 1000 is referred to as `decisive' in general scientist's nomenclature \cite{jeffreys, Jarosz2014}.
Yet in forensic science we call it `moderately strong' \cite{afsp, enfsi}, and in DNA analysis this number may even be considered too low to report \cite{Hahn2023}.
Clearly, more research is needed in how  `mid range' LRs are actually interpreted, and how we can improve this. 

\section*{Acknowledgements}
My sincere thanks to Marjan Sjerps, Wauter Bosma, David vd Vloed and Charles Berger for critical reading of this manuscript.

\newpage
\bibliographystyle{unsrt}
\bibliography{bibliography}
\newpage

\appendix

\begin{table}[]
    \centering
    \begin{tabular}{l|c}
    \hline
    certainty of bloodstain pattern type classification & LR\\
    \hline
    identification (`definitive')  & 6 \\
    possible (`included') & 1 \\
    excluded & 1 / 4 \\
    \hline

    \end{tabular}
    \caption{Likelihood ratios for bloodstain pattern evaluation conclusions by expert witnesses.\cite{bloodstain}}
    \label{tab:bloodstain_results}
\end{table}

\begin{table}[]
    \centering
    \begin{tabular}{l|c}
    \hline
    & LR\\
    \hline
    The questioned sample was written by the known writer  & 17 \\
    The questioned sample was probably written by the known writer & 7 \\
    No conclusion & 1 / 2 \\
    The questioned sample was probably not written by the known writer & 1 / 22 \\
    The questioned sample was not written by the known writer & 1 / 28 \\
    \hline

    \end{tabular}
    \caption{Likelihood ratios for handwriting evaluation conclusions by expert witnesses.\cite{handwriting}}
    \label{tab:handwriting_results}
\end{table}

\begin{table}[]
    \centering
    \begin{tabular}{l|c}
    \hline
    & LR\\
    \hline
    identification  & 113 \\
    high association & 13 \\
    association & 1  \\
    limited association & 1 \\
    no association & 1 / 6 \\
    exclusion & 1 / 7\\

    inconclusive & 1 / 2 \\
    not suitable & 1 / 2 \\
    \hline

    \end{tabular}
    \caption{Likelihood ratios for footwear print evaluation conclusions by expert witnesses.\cite{footwear}}
    \label{tab:footwear_results}
\end{table}

\begin{table}[]
    \centering
    \begin{tabular}{l|c}
    \hline
    & LR\\
    \hline
    identification  & 81 \\
    inconclusive A & 2 \\
    inconclusive B & 1 / 2 \\
    inconclusive C & 1 / 14 \\
    elimination & 1 / 28 \\
    other & 1 \\
    \hline

    \end{tabular}
    \caption{Likelihood ratios for cartridge case evaluation conclusions by expert witnesses.\cite{bullets}}
    \label{tab:cartridge_results}
\end{table}

\begin{table}[]
    \centering
    \begin{tabular}{l|c}
    \hline
    & LR\\
    \hline
    individualisation  & 376 \\
    inconclusive & 2 \\
    exclusion & 1 / 11 \\
    \hline

    \end{tabular}
    \caption{Likelihood ratios for latent fingerprint evaluation conclusions by expert witnesses.\cite{fingerprint}}
    \label{tab:fingerprint_results}
\end{table}

\end{document}